\documentclass[a4paper,12pt]{article}

\usepackage[utf8]{inputenc}
\usepackage[T1]{fontenc}

\usepackage[english]{babel}

\usepackage[numbers]{natbib}

\usepackage{amsthm,amscd,graphicx}

\usepackage{amsmath,amssymb,amsfonts,stmaryrd}

\usepackage{bbm}
\usepackage{tikz}

\usepackage{authblk}

\usepackage{color}
\usepackage{colortbl}

\usepackage[colorlinks,
            pdftex,
            pdfauthor={},
            pdftitle={asymptotics CV},
            pdfsubject={},
            pdfkeywords={}]{hyperref}

\newtheorem{theorem}{Theorem}[section]

\newtheorem{remark}[theorem]{Remark}

\theoremstyle{definition}

\DeclareMathOperator{\argmax}{argmax}

\newcommand{\ind}{\mathbbm{1}}

\newcommand{\ensnombre}[1]{\mathbb{#1}}
\newcommand{\N}{\ensnombre{N}}

\newcommand{\R}{\ensnombre{R}}

\def \E{\mathbb{E}}

\DeclareMathOperator{\Cov}{Cov}

\newcommand{\beq} {\begin{eqnarray*}}
\newcommand{\eeq} {\end{eqnarray*}}
\newcommand{\bei}{\begin{itemize}}
\newcommand{\eei}{\end{itemize}}

\usetikzlibrary{arrows.meta}
\tikzset{%
  >={Latex[width=2mm,length=2mm]},
            base/.style = {rectangle, rounded corners, draw=black,
                           minimum width=2cm, minimum height=1cm,
                           text centered, font=\sffamily},
            base2/.style = {ellipse, draw=black,
                           minimum width=4cm, minimum height=1cm,
                           text centered, font=\sffamily},
           metamodel/.style = {diamond, draw=black, thick, fill=blue!20,
                        text width=5em, text badly centered,
                        inner sep=1pt, rounded corners},
  activityStarts/.style = {base, fill=blue!30},
       startstop/.style = {base, fill=red!30},
    activityRuns/.style = {base, fill=green!30},
    activityRuns2/.style = {base, fill=gray!20},         
    process/.style = {base, minimum width=2.5cm, fill=orange!20,
                           font=\ttfamily},
         user/.style = {base2, minimum width=2cm, fill=orange!20,
                           font=\ttfamily},
         meta/.style = {metamodel, minimum width=2.5cm, fill=red,
                           font=\ttfamily},
         meta2/.style = {metamodel, minimum width=2.5cm, fill=gray!20,
                           font=\ttfamily},
         sortie/.style = {base2, minimum width=2.5cm, fill=yellow!20,
                           font=\ttfamily},
         sortie2/.style = {base2, minimum width=2.5cm, fill=gray!20,
                           font=\ttfamily},}

\usetikzlibrary{shapes}

\usepackage{a4wide}

\title{Expected Improvement applied to an industrial context - Prediction of new geometries increasing the efficiency of fans}

\author[1]{B. Demory}
\author[1]{M. Henner}
\author[2]{A. Lagnoux}
\author[3]{T.M.N. Nguyen}

\affil[1]{Valeo. 8, rue Louis Lormand
CS 80517 La Verri\`ere, France.}
\affil[2]{Institut de Math\'ematiques de Toulouse; UMR5219. Universit\'e de Toulouse; CNRS. UT2J, F-31058 Toulouse, France.}
\affil[3]{Ho Chi Minh City University of Science, 227 Nguyen Van Cu, Phuong 4, Ho Chi Minh, Vietnam.}

\begin{document}

\maketitle

\begin{abstract}
This study has been done in cooperation with the automotive supplier Valeo. In automotive industry, client needs evolve quickly in a competitiveness context, particularly, regarding the fan involved in the engine cooling module. The practitioners are asked to propose ``optimal'' new fans in short times. Unfortunately, each evaluation of the underlying computer code may be expensive whence the need of approximated models and specific, parsimonious, and efficient global optimization strategies. In this paper, we propose to use the Kriging interpolation combined with the expected improvement algorithm to provide new fan designs with high performances in terms of efficiency. As far as we know, such a use of Kriging interpolation together with the expected improvement methodology is unique in an industrial context and provide really promising results. 
\end{abstract}

\medskip


%

\section{Introduction and Motivations}
\label{sec:intro}

Many mathematical models encountered in applied sciences involve a computer code (also called a ``black-box'' simulator) given by an unknown deterministic real-valued function $f\colon D\subset \R^d\to \R$ defining an input/output relation. In several engineering problems, the goal is then to optimize the function $f$.   
In practice, the number of function evaluations may be severely limited by time or cost and the practitioners typically dispose of a very limited evaluation budget.
Consequently, the computational time required for each evaluation of the computer code together with the possibly high dimension of the input space generally do not allow an exhaustive exploration of the input space  
 under realistic industrial time constraints. 
 Moreover, in most cases, the non-availability of
derivatives prevents one from using gradient-based techniques. Similarly, the use of metaheuristics
(e.g., genetic algorithms) is compromised by severely limited evaluation budgets. Hence, such limitations pose a serious challenge to the field of global optimization and statistical approaches are mandatory to propose surrogate models and to search optima in reasonable short time. 

A first step in that direction consists in proposing mathematical approximations of the input/output relation, namely  ``metamodels'' or ``surrogate models''. These response surfaces can then be used for visualization, prediction, and optimization. 
Their construction relies on a prior knowledge consisting in available observations (data collected by evaluating the
objective function $f$ at a few points) provided by the practitioner. More precisely, the user dispose of a sample of $N$ observations $f(x_1)$, ..., $f(x_N)$ at locations $x_1$, ..., $x_N$ to realize inference and to construct an accurate metamodel. In such a framework, it is worth noticing that the uncertainty does not refer to a random phenomenon but to a partially observed deterministic one.
Due to the limited evaluations budget, the need to select cautiously evaluation points when attempting to solve this problem appears to be crucial. 
Several strategies have been developed like moving average \cite[p.36]{Ripley81}, linear regression \cite[p.29]{Ripley81}, splines \cite[p.181]{Cressie93}, Kriging interpolation \cite{Stein99, Rasmussen06}, bayesian strategies \cite{gaudard1999bayesian}, neural networks \cite{bryan2002three}... See also \cite{arnaud2000estimation,baillargeon2005krigeage} for a more complete review. 
In this paper, we consider the Kriging interpolation as metamodel. In Kriging, the unknown computer code $f$ that is to be estimated is assumed to be the realization of a Gaussian process. The exploitation of a $N$ sample $f(x_1)$, ..., $f(x_N)$ of observations at locations $x_1$, ..., $x_N$ allows working on the conditioned process which is known to be Gaussian
at any point with known mean and known variance. 
This conditional Gaussianity, together with the explicit expressions of the conditional
moments, is one of the main reasons why Gaussian processes and Kriging are attractive and have became so popular during the last decades. 

The second step wants to make use of the Kriging interpolation to proceed to the global optimization (say, maximization). The key to using response
surfaces for global optimization lies in balancing the need to exploit the approximating surface (by
sampling where it is maximized) with the need to improve the approximation (by sampling where
prediction error may be high). Proceeding to the direct optimization of the Kriging mean then appears to be appealing. Nevertheless, optimizing directly a deterministic metamodel (like the Kriging mean, or even a spline or a polynomial)  
may be inefficient and may lead to artificial optima, as shown numerically in \cite{jones2001taxonomy}. 
Fortunately, several efficient criteria have been introduced to tackle such a problem, like expected improvement, knowledge gradient,... \cite{JSW98}.
By its nice properties and its analytical tractability, expected improvement has become one of the most attractive procedure. Its principle is simple and natural: it measures the improvement brought by a point in the maximization of the function $f$ and then chooses new points that maximizes the improvement. A balance is then made between the exploration of regions where prediction error may be high and the exploitation of promising regions that would yield a notably improvement. 
 
This paper is the result of a fruitful cooperation between academic researchers of the Institut de Math\'ematiques de Toulouse and Valeo, an industrial partner. It has been realized during the project PEPITO supported by the French National
Research Agency (ANR). Aims were to experiment an extreme approach based on intensive and multiphysical simulations, the use of parametrized geometries, the determination of designs of experiments with large number of factors and the search for optima in large and high dimensional domains. 
In particular, we were interested in the fan involved in the engine cooling module that plays a key role in the engine durability and remains an important focus of the engineering teams. The practitioners are asked to propose in short time new fan designs answering the client requirements in terms of efficiency, torque, acoustics, packaging... Unfortunately, each evaluation of the computer code is time-consuming (about 3000 CPU.hour) and such a goal costly to achieve. In that view, we combined the use of the Kriging interpolation and the expected improvement algorithm to determine new ``optimal'' fan designs (with high performances). As far as we know, such an use of Kriging together with expected improvement is unprecedented in an industrial context and provide really promising results.

The paper is organized as follows. In Section \ref{sec:indus}, we present the industrial context together with the description of the input and output variables involved in the computer code. Sections \ref{sec:kriging} and \ref{sec:EI} are devoted to the presentation of the Kriging interpolation and the expected improvement optimization algorithm. The numerical results are presented in Section \ref{sec:results}. Finally, Section \ref{sec:concl} concludes this article.

\section{The industrial context}
\label{sec:indus}

Due to the ever changing geometry and architecture of cars, original equipment makers are constantly requesting new fans that perfectly fit to their needs. The specifications are given most of the time by the performances at a \textit{design point} (pressure and maximum efficiency are targeted), an \textit{off-design condition} (lower pressure but higher flow rate), and an acoustic level for the nominal point (fan noise is mostly perceptible at vehicle idle). Electrical consumption and packaging are of course part of the equation, and the resulting design must be seen as a compromise between different objectives.

\begin{figure}[!h]
\centerline{\includegraphics[width=10cm]{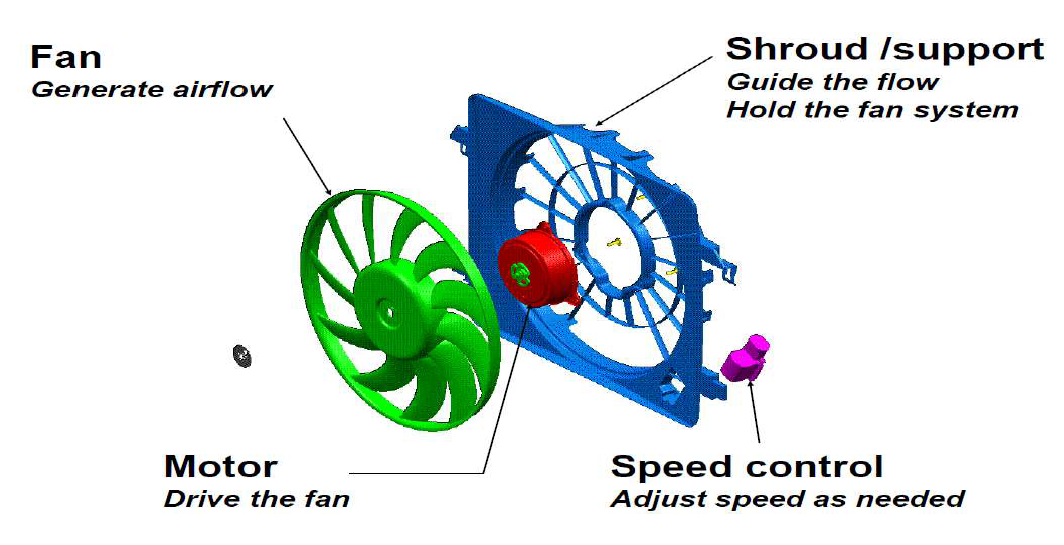}}
\caption{Fan system architecture}\label{fig:architecture}
\end{figure}

The study presented in this paper focuses particularly on the optimization of the fan blade. A representation of a typical fan system architecture is presented in Figure \ref{fig:architecture}. It shows a complex system which design requires a good expertise in both fields of turbomachine and automotive integration. 

\begin{figure}[!h]
\centerline{\includegraphics[width=10cm]{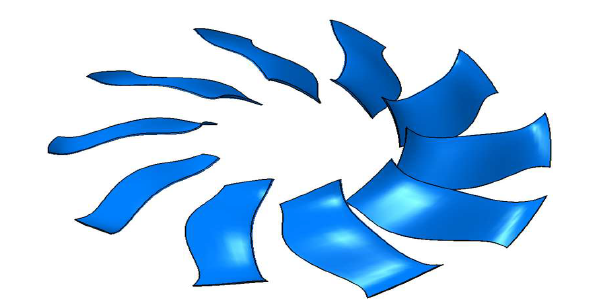}}
\caption{Innovative blade shape: not intuitive, not given by current theory}\label{fig:design}
\end{figure}

Recent developments have been greatly accelerated with the use of numerical simulation, which has allowed engineers to reduce the number of prototypes and the test campaigns. The main difficulty lies in the increased complexity which is due to the antagonist criteria of higher performance requirements and reduced space in engine compartment. In addition, the time allocated by the manufacturers to answer to any new specification with a new development is drastically reduced. In this context, the lead time of the simulation and the amount of data produced are not necessarily compatible with the multiple iterations required. Fortunately, recent advances in simulation (both hardware and software) make its use more affordable and allow the practitioner to run automated calculations daily instead of using much human time of an expert engineer. Such an opportunity opens the door to optimization processes that use simulation intensively. Interesting results have already been found as demonstrated in Figure \ref{fig:design}, with blade shapes that are innovative and non-intuitive for the expert knowledge.

In the next two subsections, we describe properly the computer code leading to the selected responses and the industrial problem.

\subsection{Description of the computer code}

The physical phenomena yielding the industrial responses like the pressure rise (downstream pressure minus upstream pressure), the torque (integral of moments due to pressure and viscous forces), the global efficiency of the fan and other local variables on both rotor and stator, accoustics, mass, size,...  
are complex. Mathematically, they are represented by an input/output relation, called the black-box model, given by
\begin{align*}
\begin{matrix}
f\colon & \R^d\times [0,\infty) & \to & \R^p\\
& (x_1,\dots,x_d,Q) & \mapsto & (y_1,\dots,y_p)=f(x_1,\dots,x_d,Q).\\
\end{matrix}
\end{align*} 
In our study, the selected responses are the pressure rise ($\Delta P$ in Pa), the torque ($C$ in N.m), and the global efficiency ($R$ in \%) of the fan so that $p=3$.
One may notice that the fan efficiency $R$ is directly related to the two other global variables $\Delta P$ and $C$ by the following relation 
\begin{align*}
R=\frac{Q\times \Delta P}{C \times \Omega},
\end{align*}
$Q$ being the flow rate (m3/s) and $\Omega$ the rotational speed (rad$^{-1}$).
In practice, the practitioner expects a static efficiency of about $55\%$.

\begin{figure}[!h]
\centerline{\includegraphics[width=10cm]{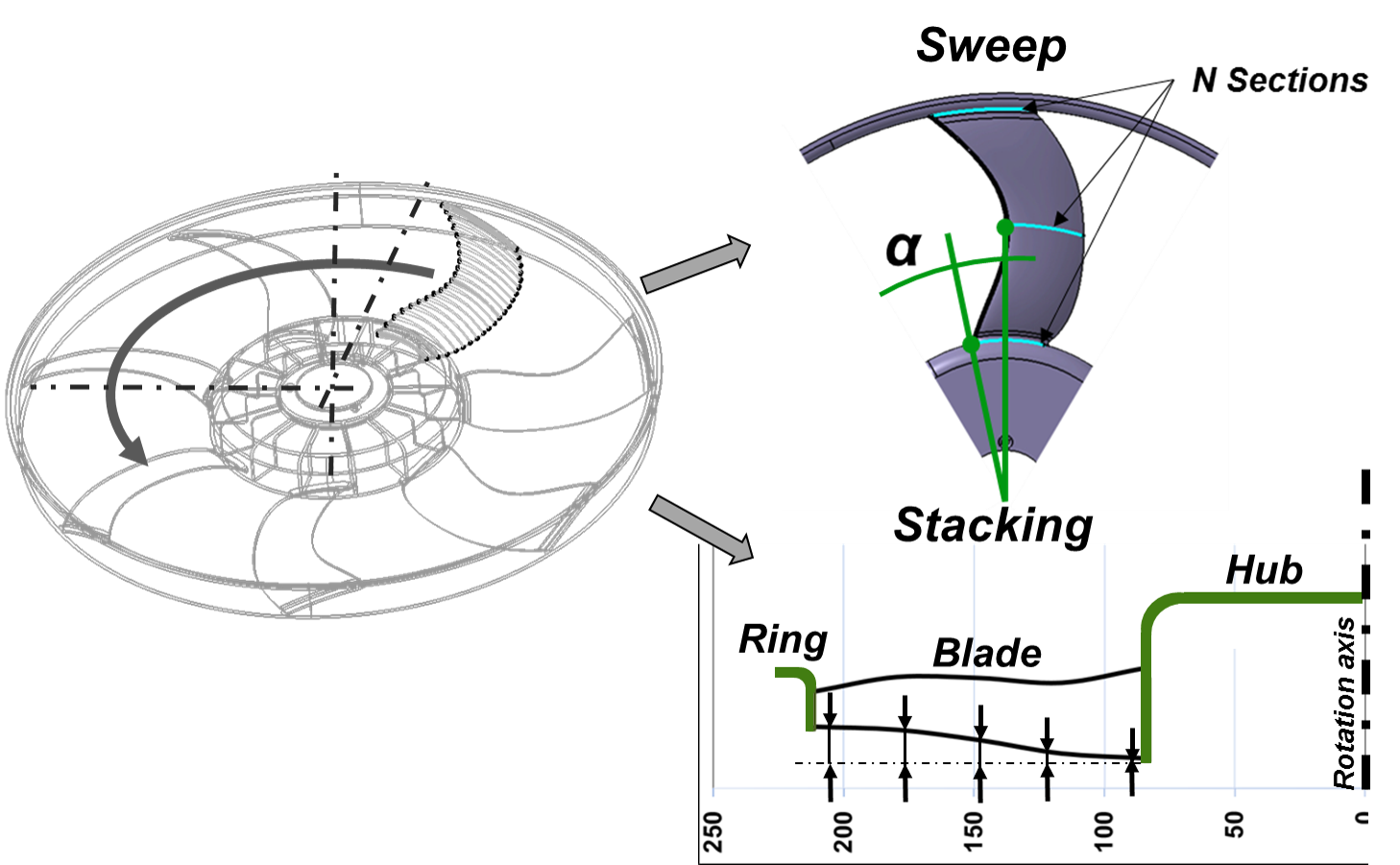}}
\caption{Fan description}\label{fig:fan}
\end{figure}

\begin{figure}[!h]
\centerline{\includegraphics[width=10cm]{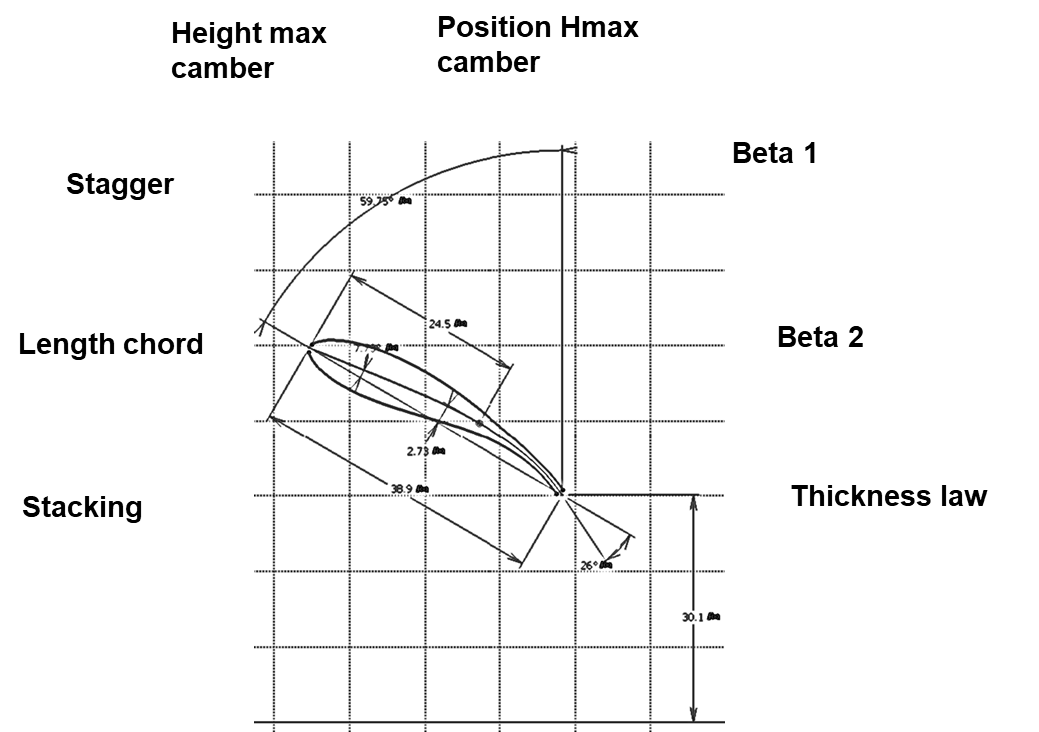}}
\caption{Profile parameters}\label{fig:fan_2}
\end{figure}

The input factors $x_1$,..., $x_d$ involved in the computer code $f$ are related to the full cooling module and in particular to the fan parametrization. 
The fan is composed of several blades which are equivalent to rotating wings. The blade section at a constant radius is an aerodynamic profile with characteristics of lift and drag, which create respectively the pressure rise and the torque of the fan. 
Some of the fan parameters are represented in Figure \ref{fig:fan} and some others specific to the blade can be found in Figure \ref{fig:fan_2}.
Using a satisfactory number of factors would lead to select 60 of them for the cooling module. Anyway, in order to make things feasible, only 14 parameters have been highlighted via a preliminary sensitivity analysis based on Sobol indices and thanks to the expert knowledge (see \cite{moreau2004geometric,grondin2005robust}), while the others are fixed to their nominal values. Twelve geometrical factors (sweep (2), max camber height (2), stagger angle (4), chord length (4)) are selected, and two others are added to size a plate behind the fan in order to represent the aerodynamic blockage due to the thermal engine of the car (it is placed behind the fan and acts as an obstacle similarly to the flow in the underhood). The flow rate is the physical factor giving the 15th parameter. Hence $d=14$.


\subsection{The industrial problem}
\label{ssec:pb}

In this study, the focus is made on innovative fan designs that will improve the fan efficiency and we want to propose an enhanced optimization method able to handle enough parameters for the design of the most efficient, quiet and compact fans in a short time frame. 

As mentioned in the introduction, direct optimization, which is based on a gradient descent algorithm, is a fairly common solution. However, it has the drawbacks of producing only local optima, and therefore of having to be repeated for each new optimization with lengthy calculation iterations. It should also be mentioned that the case of multi-objective optimization requires the creation of a cost function which sets the trade-off between the different objectives, and does not produce a Pareto front that could be analyzed by engineers. On the other hand, the exhaustive search for an optimum is made very difficult by the large number of parameters, which moreover have interactions between them. Exploration in a field of dimension 14 greatly exceeds the capabilities of an engineer, as expert as he could be in turbomachine field. Such an exploration is also intractable statistically speaking. Last but not least, each evaluation of the output is a time-consuming with traditional iterative processes and costly task. In such a context, a proper strategy for the optimization process both efficient and parsimonious is required. 

\medskip

In this paper, we chose the Kriging interpolation to construct a surrogate model easy and cheap to evaluate. Then we combine it to the so-called expected improvement optimization algorithm to build efficient new fan designs. In the next two sections, we present the Kriging interpolation also used in the sequel together with the expected improvement methodology and the description of of the numerical application and with its results are presented in Section \ref{sec:results}.

\section{Kriging}
\label{sec:kriging}

Originally introduced in geosciences by \cite{Krige51}, Kriging is a stochastic method of interpolation. The aim is to predict the value of a natural phenomenon at any arbitrary location of interest from the measured observations
at the sample points. 
 The theoretical basis was first  developed in the 1960's by \cite{Matheron62,Matheron63}. See also the famous and well-documented references on the topic \cite{Stein99,Santner03,Rasmussen06}. Nowadays, Kriging is widely used in the domain of spatial analysis and computer experiments.
From a mathematical point of view, we consider a function $f \colon x\in D\subset \R^d \mapsto f(x) \in \R$ and we wish to predict $f(x_0)$ from a sample of $N$ observations $(f(x_1),\ldots,f(x_N))$ at locations $x_1,\ldots,x_N$. 
The key ingredient of Kriging is that $f$ is assumed to be a realization of a process $Y\colon D\subset \R^d \to \R$ with mean function $m\colon x\in\R^d\mapsto m(x)\in\R$ and covariance kernel $k\colon (x,y)\in\R^d\times \R^d \mapsto k(x,y)\in\R$. 
 Then, Kriging uses a weighted average of the observations as estimate. The weights are chosen so that the Kriging prediction is unbiased with minimal variance error. 
%

In order to illustrate Kriging, we develop the methodology in the particular setting 
of what is called the \textit{simple Kriging} in which the process $Y$ is assumed to be centered and stationary at order 2 with known covariance function $k$. 
Let the vector $r_N(x_0)$ be given by
\[
r_N(x_0)=(\Cov(Y(x_0),Y(x_1)), \ldots, \Cov(Y(x_0),Y(x_N)))^\top=(k(x_0,x_1),\ldots,k(x_0,x_N))^\top
\]
and the square matrix $R_N$ of the covariances on the observation points of size $N\times N$ given by
\[
R_N=\begin{pmatrix}
\Cov(Y(x_1),Y(x_1)) & \ldots & \Cov(Y(x_1),Y(x_N))\\
\vdots & \ldots & \vdots\\
\Cov(Y(x_N),Y(x_1)) & \ldots & \Cov(Y(x_N),Y(x_N))\\
\end{pmatrix}
=
\begin{pmatrix}
k(x_1,x_1) & \ldots & k(x_1,x_N)\\
\vdots & \ldots & \vdots\\
k(x_N,x_1) & \ldots & k(x_N,x_N)\\
\end{pmatrix}.
\]

Then the (random) Kriging prediction writes
\begin{align}\label{eq:mean_kriging}
\widehat Y(x_0)=\sum_{i=1}^N \lambda_i^*(x_0) Y(x_i),
\end{align}
where the optimal vector 
\[
\lambda^*(x_0)=(\lambda^*_1(x_0),\ldots,\lambda^*_N(x_0))^\top=R_N ^{-1}r_N(x_0)
\] 
is obtained by minimizing the quadratic error.

One of the main interest of Kriging is that the Kriging variance is explicitly known: 
\begin{align}\label{eq:variance_kriging}
\sigma_N^2(x_0)=\E\left[\left(Y(x_0)-\hat Y(x_0)\right)^2\right]=k(x_0,x_0)-r_N^{\top}(x_0)R_N^{-1}r_N(x_0),
\end{align}
allowing the practitioner to build confidence intervals. Furthermore, if the underlying process $Y$ is Gaussian and observed at a given $N$-uplet $(y_1,\dots,y_N)^\top$, we have the much stronger result that the conditional distribution of $Y(x_0)$ given $(Y(x_1),\dots,Y(x_N))=(y_1,\dots,y_N)$ is Gaussian with mean $\sum_{i=1}^N \lambda_i^*(x_0) y_i$ and variance $\sigma_N^2(x_0)$
so that the Kriging prediction is the best predictor (in terms of minimizing the variance of the prediction error), linear or non linear \cite[p.140]{Rice06}. 

Notice that if covariance parameters are unknown, several estimation procedures exist; then the estimated covariance parameters are plugged in \eqref{eq:mean_kriging} and \eqref{eq:variance_kriging}. For instance, the unknown parameters can be estimated by maximum likelihood. As proposed in the function ``km'' of the R package \texttt{DiceKriging} that performs Kriging, penalized maximum likelihood estimation is also possible if some
penalty is given, or Leave-One-Out for noise-free observations.

\begin{figure}[!tpb]
\centerline{\includegraphics[width=10cm, trim = 3cm 4cm 1cm 4cm, clip]
{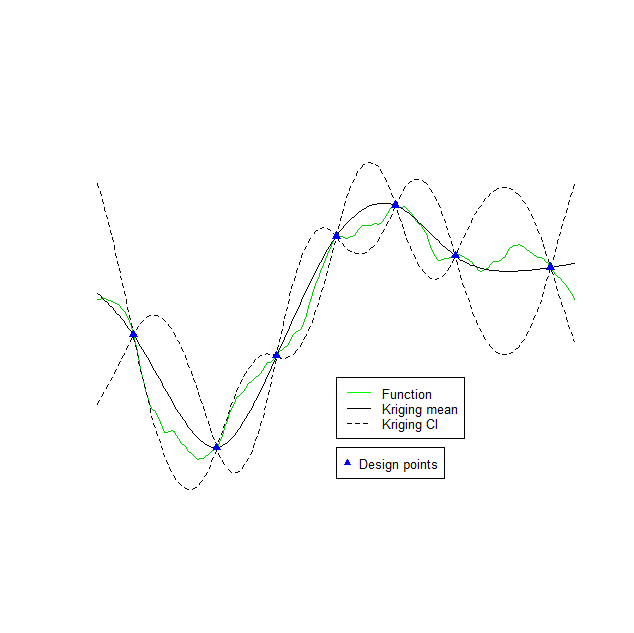}}
\caption{Example of one-dimensional data interpolation by Kriging, with confidence intervals. The green curve shows the function $f$. Triangles in blue indicate the location of the data. The Kriging interpolation, shown in black, runs along the means of the normally distributed confidence intervals shown in dashed lines. }\label{fig:Kriging}
\end{figure}

Figure \ref{fig:Kriging} represents the Kriging interpolation on a toy example. In this example, the unknown function $f$ is represented by the green curve. The practitioner provides an initial design consisting in 7
points $y_1,\ldots, y_7$ observed at $x_1,\ldots, x_7$ and represented by blue triangles. The Kriging interpolation is represented by the
black curve. The dashed curves represent the Kriging confidence intervals.

\section{Expected improvement}
\label{sec:EI}

The goal is to optimize a function $f\colon D\subset \R^d \to \R$. 
As explained in the introduction, it would be tempting to replace the costly simulator $f$ by the Kriging interpolation and to directly optimize it. Anyway, such a procedure is generally not efficient as demonstrated numerically in \cite{jones2001taxonomy}. Furthermore, it may  potentially lead to artificial optima in case of iterated optimizations with metamodel update. Fortunately, efficient criteria like the expected improvement 
 have been proposed for sequential Kriging-based optimization (see, e.g., for a comparative criteria study \cite{schonlau1997computer} and \cite{sasena2002exploration}).

In this section, we present briefly the expected improvement optimization algorithm,
 first introduced by Mockus \cite{Mockus75} and further combined with the Gaussian processes model in the efficient global optimization (EGO), see e.g. \cite{JSW98}, and sequential Kriging optimization (SKO), see e.g. \cite{HANM06}. These two methods are the key ingredient for most Bayesian optimization algorithms. The reader may follow the references therein for more details.

\paragraph{Presentation of the algorithm}

In view of maximizing the function $f$, we aim at proposing a point $x^*$ so that 
$f(x^*)$ is as close as possible to $\max_{x\in D} f(x)$.
Similarly to Kriging, we consider that $f$ is a realization of a Gaussian process $Y$ with known mean function $m$ and covariance kernel $k$.  The point $x^*$ is derived as the best point from the sample pairs $\{(x_1,f(x_1)),\ldots, (x_N,f(x_{N_s}))\}$. The points $x_1,\ldots, x_{N_s}$ are chosen repeatedly following a three steps procedure. 

\medskip

\textit{Step 1 - Initial design}. For $N$ such that ($N<N_s$) (e.g. $N=N_s/2$), we choose the points $x_1,\ldots, x_{N}$ using a space filling criterion. For instance, one may use a latin hypercube sampling (LHS) \cite{JCS05} or an orthogonal array (OA) \cite{Owen92}. Then we evaluate  $f(x_1),\ldots, f(x_N)$.

\medskip

\textit{Step 2 - Sequential incrementation}. For $n=N,\dots,N_s-1$, we derive $x_{n+1}$ from the current sample pairs $\{(x_1,f(x_1)),\dots, (x_n,f(x_n))\}$ using the distribution of $Y$ conditioned on $\{Y(x_1)=f(x_1),\dots,Y(x_n)=f(x_n)\}$. 
More precisely, the next point $x_{n+1}^{EI}$ is chosen such that
\[
x_{n+1}\in \underset{x\in D}{\argmax} ~~\E\left[\left(Y(x)-\max\{f(x_1),\dots, f(x_n)\}\right)^+\vert Y(x_1)=f(x_1),\dots,Y(x_n)=f(x_n)\right]
\]
where $(\cdot{} )^+$ stands for the positive part, namely $\max\{\cdot{},0\}$. Let us denote by $EI_n(x)$ the \textit{Expected Improvement} given by
\[
EI_n(x)=\E\left[\left(Y(x)-\max\{f(x_1),\dots, f(x_n)\}\right)^+\vert Y(x_1)=f(x_1),\dots,Y(x_n)=f(x_n)\right].
\]
Finally, we evaluate $f(x_{n+1})$. 

\medskip

Then one can proceed to the final step.
 
\medskip

\textit{Step 3 - Proposition of a new design point}. The solution is the point $x^*$ such that 
\[
x^* =\underset{x\in \{x_1,\dots,x_{N_s}\}}{\argmax} f(x).
\]

\paragraph{Expected improvement properties}

The principle of the expected improvement procedure is simple and natural: it measures the improvement brought by a point $x$ in the maximization of the function $f$ and then chooses new points that maximizes the improvement. 

The criterion $EI_n$ has nice properties: first, it is strictly positive as soon as the Kriging variance is, second, it cancels if the Kriging variance is zero and the Kriging mean is smaller than the actual maximum given by $M_n= \max\{f(x_1),\dots, f(x_n)\}$ and finally, it  increases with the Kriging mean. 
Moreover, $EI_n(x)$ has an explicit expression given by
\[
EI_n(x)=\left(\hat Y(x)-M_n \right) \Phi\left(\frac{\hat Y(x)-M_n}{\sigma_n(x)}\right)+\sigma_n(x)\phi\left(\frac{\hat Y(x)-M_n}{\sigma_n(x)}\right)
\]
where $\phi$ and $\Phi$ are respectively the probability density function and the cumulative distribution function of the standard Gaussian law and $\hat Y(x)$ and $\sigma_n^2(x)$ are respectively the Kriging mean and the Kriging variance after $n$ measurements. See \eqref{eq:mean_kriging} and \eqref{eq:variance_kriging} in Section \ref{sec:kriging} for their expressions. Consequently, one may calculate exactly $EI_n(x)$ in $O(n^2)$. More precisely, once the inversion of $R_n$  has been done (with a computational cost in $O(n^3)$) and stored, each evaluation $EI_n(x)$ of the expected improvement criterion is in $O(n^2)$ due to the computation of the quadratic form $r_n(x)^\top R_n^{-1}r_n(x)$ involved in the conditional variance $\sigma_n^2(x)$ (see \eqref{eq:variance_kriging} in Section \ref{sec:kriging}). Convergence guarantees for the expected improvement algorithm are given in  \cite{VB10,BBG19}. 

Another methodologies for choosing $x_{n+1}$ have been developed like knowledge gradient that is a close variant of the expected improvement algorithm \cite{FPD08,SFP11}. Anyway, the exact computation of the knowledge gradient function being more costly than expected improvement, practitioners prefer to use the expected improvement algorithm.

\begin{figure} 
\begin{tabular}{c}
\includegraphics[width=12cm, trim = 1cm 0cm 1cm 0cm, clip]{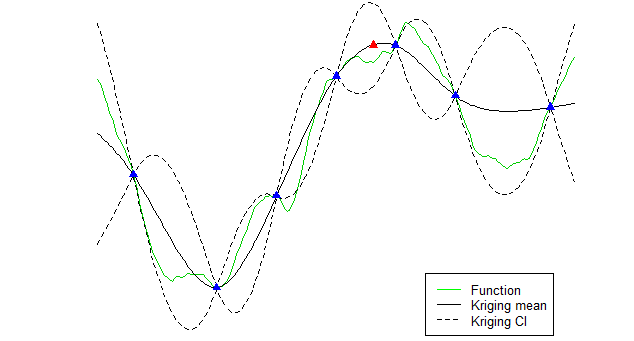}\\
\includegraphics[width=12cm, trim = 1cm 0cm 1cm 0cm, clip]{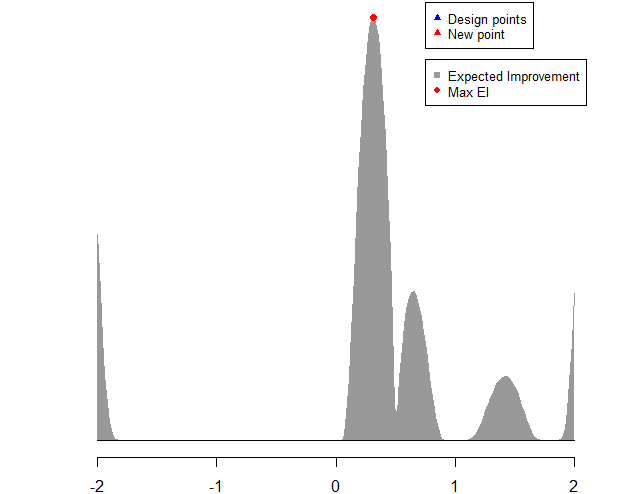}\\
\end{tabular}
\caption{Illustration of the expected improvement algorithm applied to Kriging}
\label{fig:Kriging_EI}
\end{figure}

Figure \ref{fig:Kriging_EI} represents the expected improvement algorithm on a toy example.
In this example, the unknown real code is represented by the green curve. The practitioner provides an initial design consisting in 7 points represented by blue triangles. First, we proceed to the Kriging interpolation leading to the black curve. The dashed curves represent the confidence intervals of the Kriging interpolation. Second, we proceed to the computation of the expected improvement criterion. Finally, the new point to predict is chosen in the most promising region: with highest value of expected improvement.

\paragraph{Parallelizations: expected improvement multi-points}

Now, the goal is to propose several new points at each iteration of the algorithm. In that view, the second step of the sequential incrementation is updated in the following way.

\medskip

\textit{New step 2 - Sequential incrementation by batch} Let $b\in \N^*$ be the size of the batch, namely the number of new points to be proposed at each iteration. 
For $n=N,\dots,N_s-1$, we derive $x_{n+1,1},\dots,x_{n+1,b}$ from the current sample pairs 
\[
\{(x_1,f(x_1)),\dots, (x_N,f(x_N)),(x_{N+1,1},f(x_{N+1,1})),\dots, (x_{N+1,b},f(x_{N+1,b})), \dots, (x_{n,b},f(x_{n,b}))\}
\]
and using the distribution of $Y$ conditioned on $\{Y(x_1)=f(x_1),\dots,Y(x_k)=f(x_k), Y(x_{k+1,1})=f(x_{k+1,1}),\dots, Y(x_{k+1,b})=f(x_{k+1,b}), \dots, Y(x_{n,b})=f(x_{n,b})\}$. To shorten notation, we still denote the observation vector $y$. 
We also introduce the current maximum:
\[
M_n= \max\{f(x_1),\dots,f(x_k), f(x_{k+1,1}),\dots, f(x_{k+1,b}), \dots, f(x_{n,b})\}.
\]
Now the expected improvement criterion rewrites as:
\[
\left(x_{n+1,1},\dots, x_{n+1,b}\right)\in \underset{(x_1',\dots,x_b')\in D^b}{\argmax} ~~\E\left[\left(\max\{Y(x_1'),\dots,Y(x_b')\}-M_n\}\right)^+\vert y\right].
\]
The quantity to optimize in the right-hand side of the previous equation is naturally denoted by $EI_n(x_1',\dots,x_b')$ . Unlike in the case of a single new point at each iteration, the evaluation of 
$EI_n(x_1',\dots,x_b')$ is now complex. In \cite{CG13}, the authors proposed different strategies to compute it.

First, one may use a Monte Carlo scheme noticing that $EI_n(x_1',\dots,x_b')$  is the expectation of the function of a Gaussian vector with a known distribution. 

Second, the heuristic Constant Liar (CL) method may also be  used \cite{GLRC08,CG13}. 
To begin, the regular expected improvement is maximized.
Then, for the next points, the expected improvement is maximized again, but with an artificially
updated Kriging model. Since the response values corresponding to the last best point obtained are
not available, the idea of CL is to replace them by an arbitrary constant value (the "lie") set by the
user. We proceed repeatedly  so that the trick relies in the fact that only single point expected improvement need to be evaluated. More precisely, for any $i=1,\dots,b$, we assume that $Y(x_{n+1,1})=\tilde y_1, \dots, Y(x_{n+1,i-1})=\tilde y_{i-1}$, we set 
\[
x_{n+1,i} \in \underset{x\in D}{\argmax}~~ \E\left[\left(Y(x)-\max\{f(x_1),\dots, f(x_{n,b}), \tilde y_1,\dots,  \tilde y_{i-1} \}\right)^+\vert y\right],
\] 
we introduce a value $\tilde y_i$, and so on. The values $\tilde y_{1},\dots, \tilde y_{b-1}$ can be chosen as the maximum of all the observed values of $f$ so that the expected improvement algorithm tends to explore the function near the current maximum (as
the lie is a high value and we are maximizing $f$). Besides,     
taking the minimum of all the observed values of $f$ leads to a more exploratory expected improvement procedure. Naturally, considering the current maximum (respectively minimum) is expected to perform well on unimodal (resp. multimodal). Alternatively, one may use the Kriging mean as liars or even mix strategies taking both the current minimum and the current maximum. Then, at each iteration, two batches are generated with both strategies. From these two candidate batches, the batch with the best actual expected improvement value is chosen.

Third, one may compute exactly $EI_n(x_1',\dots,x_b')$ via $b^2$ evaluations of  multidimensional Gaussian cumulative distribution functions.

\paragraph{Extension to multi-objectives}

Similarly to co-Kriging, the expected improvement algorithm  can be generalized to optimize a multivariate function $f=(f_1,\dots,f_p)$ from 
$D\subset R^d$ to $\R^p$. In that view, the several objectives $f_1,\dots,f_p$ are considered as realizations of $p$ Gaussian processes $Y_1,\dots,Y_p$. The conditional expectation now represents the expectation conditioned on all the observed values $\{f_1(x_1),\ldots,f_1(x_n),f_2(x_1),\ldots,f_p(x_n)\}$. The optimization is seen as a sequential reduction of the
volume of the excursion sets below the current best solutions and the strategy chooses the points that
give the highest expected reduction. The reader may refer to \cite{Picheny15} for the details of such a generalization.

\section{Prediction of new geometries}
\label{sec:results}

\subsection{Available data and software}\label{ssec:soft_data}

An automated simulation process has been implemented to drive design of experiment plans, using several softwares that are commonly used in the industry.  At first, the design of the fan system has been completely parameterized in a computer-aided design (CAD) tool named Catia 
\footnote{https://www.3ds.com/products-services/catia/}, 
according to rules that allow all combinations of geometric parameters in their possible ranges of variation, while ensuring their independence.
CAD files are exported in a standard format and re-read by a fluid simulation software (StarCCM+), which provides by scripts automated meshes, model solving and automatic post-processing of results.
A third tool used for optimization (Isight\footnote{https://www.3ds.com/products-services/simulia/products/isight-simulia-execution-engine/latest-release/}) ensures the sequence of tasks by imposing the set of parameters and launching CAD and simulation tools. The different sets of data are given by Latin hypercube sampling (LHS) plans (see, e.g.\, \cite{JCS05} for an introduction to LHS) produced by Isight 
 or by factorial plans\footnote{A great care was given to the simulation convergence, all runs being maintained until monitored performances were stabilized and residuals went below strict criteria. In addition, some runs were repeated on different machines or with different rules of parallelization to check that no discrepancy appears in the process.}.

Once the quality of the simulations is proved and the results are obtained for the various plans, some initial meta-models have been produced by neural networks (Isight, Radial basis function models) and their accuracies checked by comparing the prediction (neural network) and the real experiment (simulation). Some good results were observed, and in general the trends are correctly predicted when moving one parameter. However the accuracy of the model is questionable since the number of runs is still very low compared to the size of the domain. Large errors are frequently observed which justifies some additional effort towards the implementation of a good optimization process which is the goal of this paper.

Concretely, Valeo engineers provide us a design of experiment together with the corresponding selected responses in order to lead our statistical study. More precisely, they supply a design of experiment of 300 geometries in $\R^{14}$ using the OLHS procedure. For any geometry, they compute  
the pressure rise $\Delta P$, the torque $C$ and the efficiency $R$
at a flow rate $Q=Q_{\textrm{DoE}}$ whose value runs between $1000$ and $4000$. It must be noted that the physical factor for the flow rate $Q$ has been set to $Q_{\textrm{DoE}}$ according to the plan, and that the simulations were done additionally twice for the two different flow rates $Q_{\textrm{low}}=1000$ and $Q_{\textrm{high}}=4000$.
 In addition, they provide 600 geometries in $\R^{14}$ also constructed via OLHS and the values of $(\Delta P,C,R)$ for any geometry at two different flow rates $Q$: $Q_{\textrm{DoE}}$ and $Q_{\textrm{high}}=4000$. In this paper, we focus on the efficiency only.

\medskip

The numerical experiments are implemented using the R packages \texttt{DiceKriging} and \texttt{DiceOptim} to perform respectively Kriging and expected improvement (see \cite{RGD12}).

\subsection{Preliminary study and selection of the Kriging interpolation}

A preliminary study has been led to compare different strategies to model the data. First, we considered a linear regression model on the efficiency $R$ given by:
\[
R_{\text{lin}}(g;Q)=\beta_0+\sum_{j=1}^{14} \beta_j g_j+\beta_{15} Q+\varepsilon
\]
where the $\beta_j$'s are unknown coefficients, $g=(g_j)_{j=1,\dots,14}$ is the input multivariate variable, namely the geometry, $Q$ is the flow  rate considered and $\varepsilon$ is a white noise.
The second model considered was given by:
\[
R_{\text{approx}}(g;Q)=\alpha_0+\sum_{j=1}^{14} f_j(g_j)+f_{15}(Q)+\varepsilon
\]
where $\alpha_0$ is an unknown constant, $(f_j)_{j=1,\dots,15}$ are deterministic approximation functions (e.g.\ natural cubic splines) and  $\varepsilon$ is a white noise.
Several simulations showed that 
 Kriging outperfoms the latter two models in terms of precision.

\subsection{Experiment description}\label{ssec:methodo}

As explained previously, the aim is to propose several fan designs leading to good performances in terms of efficiency. In that view, we exploit the two designs of experiment provided by Valeo and presented in Section \ref{ssec:soft_data}. Then we combine Kriging and expected improvement to achieve this goal.

\paragraph{Kriging and expected improvement on the efficiency $R$ at $Q_{\textrm{high}}=4000$ $m^3/h$}

In this section, we consider all the available information gathering the data available corresponding to a flow rate $Q_{\textrm{high}}$ equal to 4000 $m^3/h$. Namely, we consider the 900 different geometries of the two designs of experiment provided by Valeo and their corresponding efficiencies at $4000$ $m^3/h$.
Then, we first perform simple Kriging on these efficiencies using the function ``km'' of the R package \texttt{DiceKriging}. In other words, as explained in Section \ref{sec:kriging}, we consider that the efficiency $R$ at $Q_{\textrm{high}}=4000$ $m^3/h$ is the realization of a Gaussian process with unknown mean $\mu$ and covariance function $k$:
\begin{align*}
R(g;4000)=\mu(4000)+\varepsilon(g;4000)
\end{align*}
for any geometry $g$ living in $\R^{14}$. In the sequel, we assume that $\varepsilon$ is a centered homoscedastic Gaussian process with variance $\sigma^2$. In addition, we use a separable covariance function $k$:
\begin{align*}
k(g,g')=\Cov(\varepsilon(g;4000),\varepsilon(g';4000)) = \sigma^2\prod_{j=1}^{14} \rho_{\theta_j}(|g_j-g'_j|)
\end{align*} 
where the $\rho_{\theta_j}$'s are unidimensional Mat\'ern kernels with parameter 5/2:
\begin{align*}
\rho_{\theta_j}(d)= \left(1+\frac{\sqrt{5}d}{\theta_j}+\frac{5d^2}{3\theta_j^2}\right)\exp\left(-\frac{\sqrt{5}d}{\theta_j}\right).
\end{align*} 
The Mat\'ern kernel has been preferred to the Gaussian kernel or to the exponential kernel since it corresponds to a trade-off between the latter two. Notice that the value of the correlation length $\theta_j$ which is unknown has great influence on the results: then it is of crucial importance. Hence,  all the unknown parameters  $\theta_j$, for $j=1,\dots,14$  have been estimated automatically by the algorithm (data-driven algorithm) using maximum likelihood estimation, so do the two others 
 unknown parameters, namely the unknown mean $\mu$ and the unknown variance $\sigma^2$ of the underlying Gaussian process.

Second, in order to maximize the efficiency at a flow rate $Q$ equal to 4000 $m^3/h$, we run the expected improvement algorithm using the function ``max$\_$qEI'' of the R package \texttt{DiceOptim} providing a batch of $n=10$ new promising geometries. The optimization is realized with a multistarted brute
force qEI maximization with Broyden-Fletcher-Goldfarb-Shanno (BFGS) algorithm that is an iterative method for solving unconstrained nonlinear optimization problems. The BFGS method belongs to quasi-Newton methods that seek a stationary point. See \cite{Fletcher87} for more details on the BFGS algorithm.

\paragraph{Kriging and expected improvement on the estimated efficiency at $Q=2500$}

Here, alternatively, we aim at working at a nominal flow rate of 2500 $m^3/h$ rather than at the maximum flow rate of 4000 $m^3/h$ previously considered.  Unfortunately, the data at $Q=2500$ $m^3/h$ are not available. Instead, in the first design of experiment of size 300, Valeo supplied us three efficiencies $R_{\textrm{DoE}}$, $R_{\textrm{low}}$ and $R_{\textrm{high}}$ associated to 300 geometries at flow rates $Q_{\textrm{DoE}}$ (whose value runs between $Q_{\textrm{low}}=1000$ $m^3/h$ and $Q_{\textrm{high}}=4000$ $m^3/h$), $Q_{\textrm{low}}=1000$ $m^3/h$ and $Q_{\textrm{high}}=4000$ $m^3/h$, respectively. 
Guided by the expert knowledge, we realize a quadratic regression on these efficiencies leading to one quadratic curve per geometry. 
More precisely, for any fixed geometry, we assume that $Q\mapsto R(g;Q)$ is a quadratic function of $Q$ and we proceed to an interpolation of the efficiency using  the efficiency values at $Q_0=0$, $Q_{\textrm{low}}=1000$ $m^3/h$, $Q_{\textrm{DoE}}$ and $Q_{\textrm{high}}=4000$ $m^3/h$ of the 300 configurations of geometrical input parameters. Hence, at a fixed geometry $g$, the function $R(g;\cdot{})$ is approximated by the quadratic function $\widetilde R(g;\cdot{})$ given by 
\[
\widetilde R(g;Q)=a(g)\cdot{}Q^2+b(g) \cdot{}Q+c(g).
\]

\begin{center}
\begin{tikzpicture}[scale=0.9]
\draw[line width=1.6pt,color=red, smooth,samples=100,domain=-2:2] plot(\x,-\x*\x);

\draw [<-]  (0.1,0) --(3.8,-1.8*1.8+3.2) ;
\draw (0,0) node {$\bullet$};
\draw (3.8,-1.8*1.8+3.2) node[right] {Maximum efficiency};

\draw (3.8,-1.8*1.8+2) node[right] {Observations};

\draw (-2,-3.84) node {$\bullet$} ;
\draw (-3.2,-3.84) node {$R(g;0)=0$} ;
\draw [<-]  (-2+0.1,-3.84) --(3.8,-1.8*1.8+2) ;

\draw (-1.6,-2.86) node {$\bullet$} ;
\draw (-2.8,-2.86) node {$R(g;Q_{\textrm{DoE}})$} ;
\draw [<-]  (-1.6+0.1,-2.86) --(3.8,-1.8*1.8+2) ;

\draw (-1.1,-1.21) node {$\bullet$} ;
\draw (-2.3,-1.21) node {$R(g;1000)$} ;
\draw [<-]  (-1,-1.21) --(3.8,-1.8*1.8+2) ;

\draw (1.2,-1.69) node {$\bullet$} ;
\draw (0.2,-1.69) node {$R(g;4000)$} ;
\draw [<-]  (1.2+0.1,-1.69) --(3.8,-1.8*1.8+2) ;
\end{tikzpicture}
\end{center}

Then, we compute the estimated efficiency $\widetilde R$ at $Q=2500$ $m^3/h$ of the 300 geometries of the design of experiment: $\widetilde R(g;2500)=a(g)\cdot{}(2500)^2+b(g)\cdot{} 2500+c(g)$. 
Finally, we perform Kriging on these estimated efficiencies $\widetilde R(g;2500)$ at $Q=2500$ $m^3/h$ and we run the expected improvement optimization algorithm to provide a batch of 10 promising fan geometries. We follow the procedure and choices of the previous paragraph. 

\begin{remark} 
Observe that we could have proceeded in a slightly different way to estimate the efficiencies at $Q=2500$ $m^3/h$ by performing a Kriging interpolation of the efficiency $R$ in $\R^{15}$, considering the flow rate as an input parameter that may vary rather than working in $\R^{14}$ with fixed flow rate. 
\end{remark}

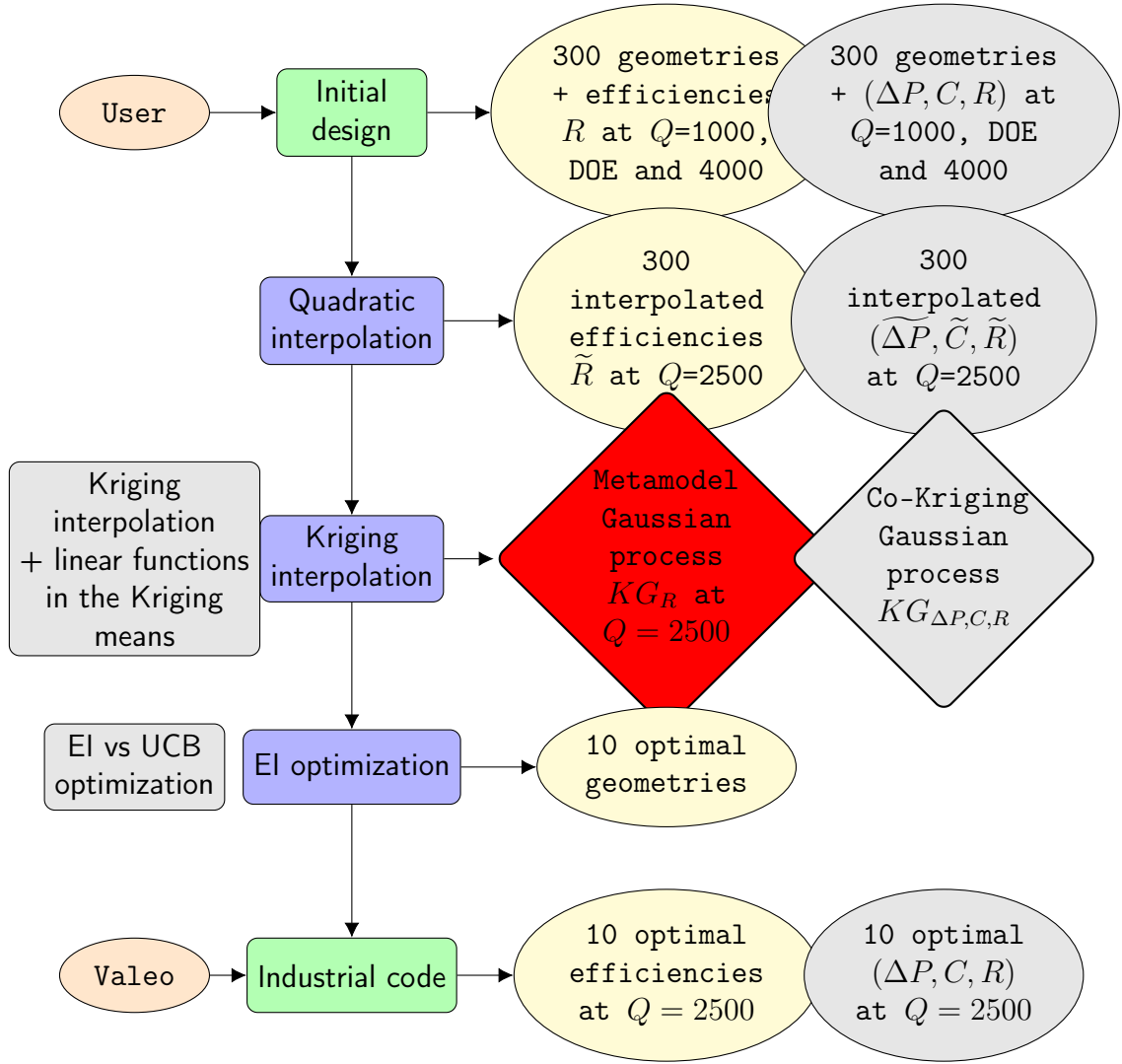
\begin{figure}[!h]
\centering
\begin{tikzpicture}[node distance=1.8cm,
    every node/.style={fill=white, font=\sffamily}, align=center]
  \node (InitialDesign)     [activityRuns]          {Initial\\ design};
  \node (Code1D)      [activityStarts, below of=InitialDesign, yshift=-1cm]   {Quadratic\\ interpolation};
  \node (Kriging)     [activityStarts, below of=Code1D, yshift=-1.4cm]   {Kriging\\ interpolation};
    \node (Optimization)      [activityStarts, below of=Kriging, yshift=-1cm]
                                                      {EI optimization};

  \node (Code1D2)      [activityRuns, below of=Optimization, yshift=-1cm]
                                                                {Industrial code};

  \node (User)    [user, right of=InitialDesign, xshift=-4.7cm]
                                                             {User};
  \node (Kriging2)     [activityRuns2, right of=Kriging, xshift=-4.7cm]   {Kriging\\ interpolation\\
  + linear functions\\ in the Kriging\\ means};

    \node (Optimization2)      [activityRuns2, right of=Optimization, xshift=-4.7cm]
                                                      {EI vs UCB\\ optimization};

  \node (User2)    [user, right of=Code1D2, xshift=-4.7cm]
                                                             {Valeo};

 \node (InitialInputs)    [sortie, left of=InitialDesign, xshift=6cm] {300 geometries\\ + efficiencies \\  $R$ at $Q$=1000,\\ DOE  and 4000};

 \node (InitialInputs2)    [sortie2, left of=InitialDesign, xshift=9.7cm] {300 geometries \\ + $(\Delta P,C,R)$ at \\ $Q$=1000, DOE\\ and 4000};

 \node (InitialOutputs)      [sortie, left of=Code1D, xshift=6cm]
                                                        {300\\ interpolated\\ efficiencies\\ $\widetilde R$ at $Q$=2500};

 \node (InitialOutputs2)      [sortie2, left of=Code1D, xshift=9.7cm]
                                                        {300\\ interpolated\\ $(\widetilde{\Delta P},\widetilde C,\widetilde R)$\\ at $Q$=2500};

 \node (Metamodel)    [meta, left of=Kriging, xshift=6cm] {Metamodel Gaussian process $KG_R$ at $Q=2500$};

 \node (Metamodel2)    [meta2, left of=Kriging, xshift=9.7cm] {Co-Kriging\\ Gaussian process $KG_{\Delta P, C,R}$};

  \node (NewInputs)      [sortie, left of=Optimization, xshift=6cm]
                                                        {10 optimal\\ geometries};

   \node (NewOutputs)      [sortie, left of=Code1D2, xshift=6cm]
                                                        {10 optimal\\ efficiencies \\ at $Q=2500$};

   \node (NewOutputs2)      [sortie2, left of=Code1D2, xshift=9.7cm]
                                                        {10 optimal\\ $(\Delta P,C,R)$\\ at $Q=2500$};

  \draw[->]     (InitialDesign) -- (Code1D);
  \draw[->]      (Code1D) -- (Kriging);
  \draw[->]     (Kriging) -- (Optimization);
  \draw[->]      (Optimization) -- (Code1D2);

    \draw[->]             (User) -- (InitialDesign);
    \draw[->]             (User2) -- (Code1D2);

    \draw[->]             (InitialDesign)  -- (InitialInputs);
    \draw[->]             (Code1D)  -- (InitialOutputs);

    \draw[->]             (Kriging)  -- (Metamodel);

    \draw[->]             (Optimization)  -- (NewInputs);
    \draw[->]             (Code1D2)  -- (NewOutputs);
\end{tikzpicture}
\caption{Procedure diagram when working at $Q=2500$ $m^3/h$. The gray boxes represent the directions for further research: adding linear functions in the Kriging means to improve the optimization step, comparing the upper confidence bound optimization procedure to expected improvement, and consider the whole vector $(\Delta P, C,R)$ of outputs and perform co-Kriging ad multi-objective expected improvement.}\label{fig:diagramme}
\end{figure}

Figure \ref{fig:diagramme} synthesizes the different steps of the procedure adopted in this article. The gray boxes represent the directions for future work: adding linear functions in the Kriging means to improve the optimization step, comparing  the upper confidence bound (UCB) optimization procedure to expected improvement, and consider the whole vector $(\Delta P, C,R)$ of outputs and perform co-Kriging ad multi-objective expected improvement. When working at $Q=4000$ $m^3/h$, a similar diagram could be drawn skipping the quadratic interpolation step. See, for instance, \cite{ACBF02} for the description of the UCB optimization methodology and \cite{SKKS12} for the theoretical asymptotic properties of this algorithm.

\paragraph{Notation} The Kriging means provided by the Kriging procedure will be denoted $\widehat R$ in both settings (considering the efficiencies at $Q_{\textrm{high}}=4000$ $m^3/h$ or considering the interpolated values of the efficiencies at $Q=2500$ $m^3/h$), while the corresponding Gaussian processes will be denoted by $KG_R$.

\subsection{Results}\label{ssec:results}

\paragraph{Performance of Kriging}

To measure the performance of the model, we realize a cross-validation procedure by Leave-One-Out (LOO). The LOO procedure consists in computing the prediction at a design point when the corresponding observation is removed from the learning set (and this, for all design points). See, for instance, \cite{Cressie93,Ripley81,Bachoc13} for more details on the LOO procedure. In that view, we use the function ``leaveOneOut.km'' of of the R package \texttt{DiceKriging} that, for any geometry $i$ of the design of experiment, determines the associated Kriging model based on the learning sample without the $i$th observation point.  The output of ``leaveOneOut.km'' consists in two vectors of length the number of observations $N$ whose $i$th coordinates correspond to the Kriging mean and the Kriging standard deviation at the $i$th observation point when removing it from the learning sample. 

\begin{figure}
\begin{tabular}{cc}
\includegraphics[width=7cm, trim=0cm 0cm 0cm 2cm, clip]{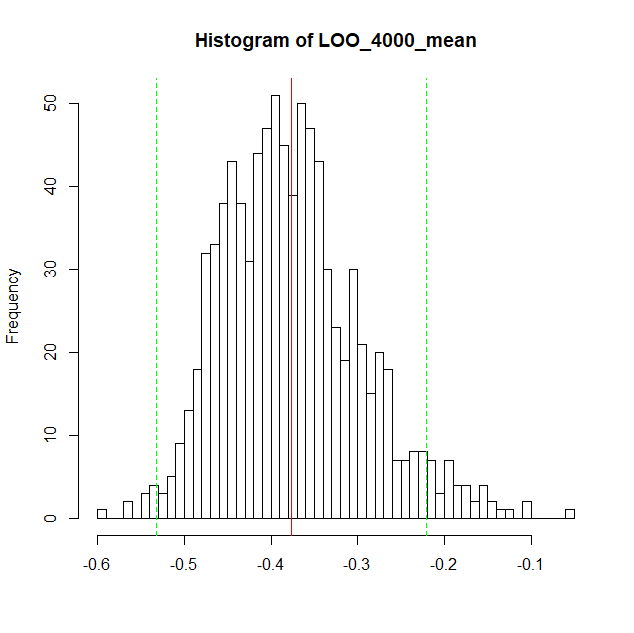}
&
\includegraphics[width=7cm, trim=0cm 0cm 0cm 2cm, clip]{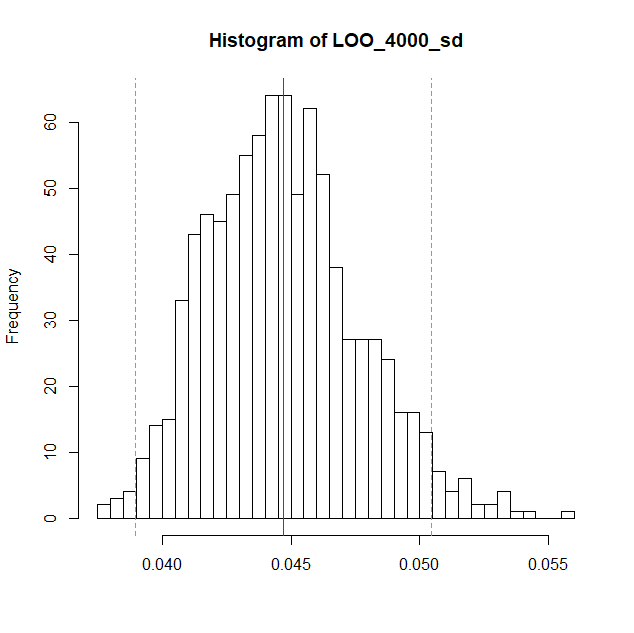}
\\
\includegraphics[width=7cm, trim=0cm 0cm 0cm 2cm, clip]{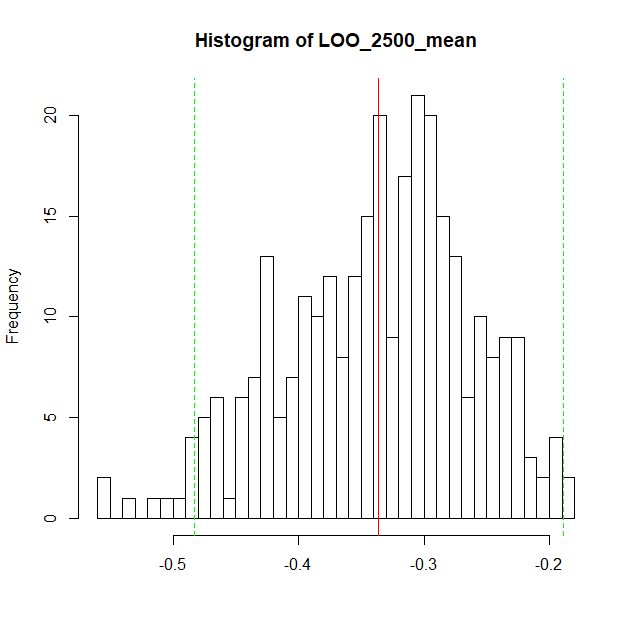}
&
\includegraphics[width=7cm, trim=0cm 0cm 0cm 2cm, clip]{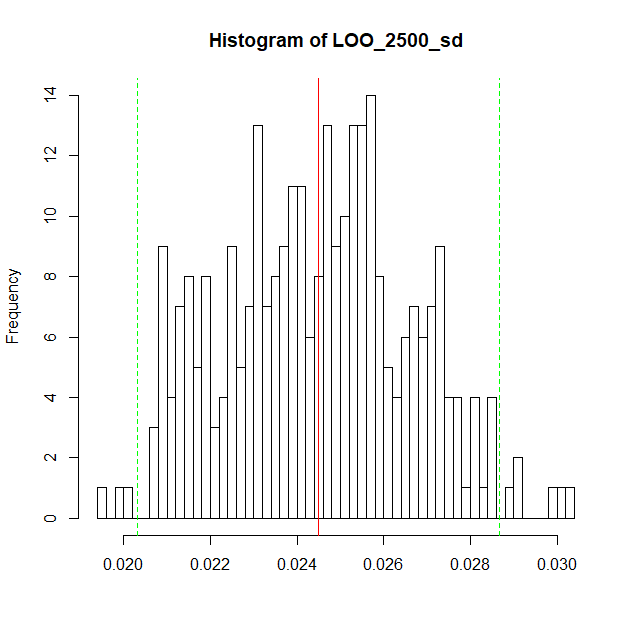}
\\
\end{tabular} \label{fig:loo}
\caption{\textbf{Kriging performances}. Histograms of the mean (left) and standard deviations (right) Leave-One-Out performed on $\widehat R(\cdot{};4000)$ (up) and $\widehat R(\cdot{};2500)$ (bottom). The vertical red line denotes the mean value. The vertical green lines corresponds to the confidence intervals at 95$\%$.}
\end{figure}

In Figure \ref{fig:loo}, we represent the histograms of these two vectors obtained considering $\widehat R(\cdot{};4000)$ and $\widehat R(\cdot{};2500)$ together with the confidence intervals at 95$\%$.  First, no outliers are highlighted in the histograms of the conditional means (left side of Figure \ref{fig:loo}). Second, we observe that the conditional standard deviations are far from constant (right side of Figure \ref{fig:loo}), traducing the fact that the observations of the design of experiment are informative. 

In Table \ref{table:Kriging_perf}, we display some classical performance criteria of the Kriging realized on the values of $\widehat R(\cdot{};4000)$ and  $\widehat R(\cdot{};2500)$. Namely, we give 
\begin{itemize}
\item[$-$] the multiple $R$-squared error ($R^2$) defined by 
\[
R^2=1-\frac{\sum_{i=1}^N(y_i-\widehat y_i)^2}{\sum_{i=1}^N(y_i-\overline y_N)^2},
\]
where $\widehat y_i$ is the prediction of the $i$-th data $y_i$ and $\overline y_N$ stands for the empirical mean of the data. The closest the value of one is, the best the prediction is;
\item[$-$] the root of the Mean Square Error (RMSE) error between the exact value and the predicted one given by
\[
\textrm{RMSE}=\sqrt{\frac1N \sum_{i=1}^N(y_i-\widehat y_i)^2};
\]
\item[$-$] the Relative Maximum Absolute (RMA) error defined by 
\[
\textrm{RMA}=\max_{i=1,\ldots,N}\frac{|y_i-\widehat y_i|}{\sigma},
\]
where $\sigma$ stands for the standard deviation of the vector of the output values;
\item the Covering Rates (CR) at $95\%$ are given by:
\[
\textrm{CR(95\%)}=\frac{1}{N}\sum_{i=1}^N \ind_{\{|y_i-\widehat y_i|\leqslant 1.96\times  \sigma_N(x_i)\}}
\]
where $\sigma_N^2(x_i)$ stands for the Kriging variance.\\
\end{itemize}

Notice that the standard deviation of the efficiencies corresponding to the flow rate $Q=4000$ $m^3/h$ (resp. $Q=2500$ $m^3/h$) is $0.092$ (resp. $0.079$).

\begin{table}[h]
\begin{center}
\begin{tabular}{c|c|c}
& Kriging on $R_{4000}$ & Kriging on $\widehat R_{2500}$\\
\hline
R$^2$ & 0.763 & 0.906\\
RMSE & 0.045 & 0.024\\
RMA & 2.281 & 0.888\\
CR(95\%) & 0.941 & 0.939\\
\end{tabular}
\end{center}
\caption{\label{table:Kriging_perf} Kriging performances on both models ($R_{4000}$ and $\widehat R_{2500}$)}
\end{table}

Both methodologies provide good results with $R^2$ values close to 1 and RMSE and RMA values close to 0 as expected. Surprisingly, the Kriging on the estimated efficiency at 2500 $m^3/h$ outperforms the Kriging at  
4000 $m^3/h$. Thus working on an average flow rate is more efficient than working at an extreme rate of 4000 $m^3/h$ and provides more accurate results. This observation and these results validate this second procedure at 2500 $m^3/h$. Observe that the covering rates appear to be under-estimated. 

\medskip

Notice that adding linear functions of parameters to the Kriging mean
would have been interesting and probably could have significantly improved the optimization. Such an improvement will be considered in a further study.


%


\paragraph{Performance of expected improvement and results}

\begin{figure} 
\begin{tabular}{cc}
\includegraphics[width=7cm, trim=0cm 0cm 0cm 2cm, clip]{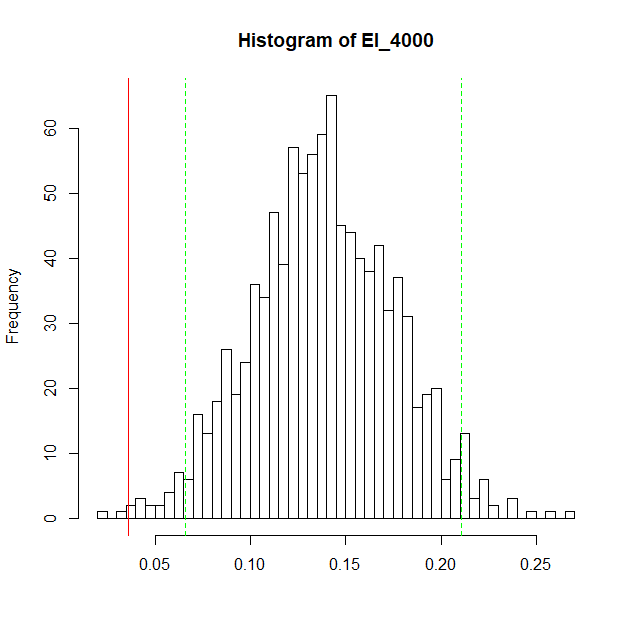}
&
\includegraphics[height=6cm, width=5.5cm, trim=0cm 0cm 0cm 2cm, clip]{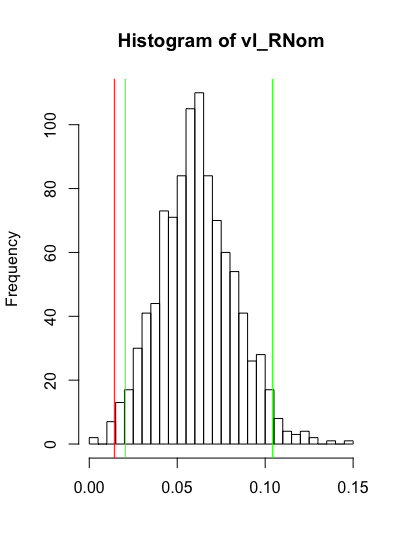}
\end{tabular}

\caption{\textbf{Expected improvement performances}. Finite sample distributions of expected improvement (histograms) computed on $\widehat R(\cdot{};4000)$ (left) and $\widehat R(\cdot{};2500)$ (right). The vertical red line denotes the a posteriori value of the EI. The vertical green lines corresponds to the confidence intervals at 95$\%$.}
\label{fig:histograms}
\end{figure}

Now, the expected improvement algorithm gives us a batch of $n=10$ new geometries $(g^{\textrm{new}}_1,\dots,g^{\textrm{new}}_{10})$, where $g^{\textrm{new}}_i\in \R^{14}$, for all $i=1,\dots, 10$. Valeo engineers then compute the corresponding efficiencies.
In order to have an idea of the likely improvements brought by expected improvement, we generate $1000$ realizations of the efficiencies $R$ at the 10 new geometries and $Q=4000$ $m^3/h$ conditionally to the data (the 300 efficiencies of the design of experiment) using the Kriging model. In other words, if $KG_R$ represents the Gaussian process obtained by the Kriging procedure from the initial design of experiment, we obtain 1000 realizations of the random vector:
\[
(KG_R(g^{\textrm{new}}_1;4000), \dots, KG_R(g^{\textrm{new}}_{10};4000)).
\] 
Then, for each of the 1000 realizations, we compute the associated value of expected improvement, namely,
\[
\Bigl( \max_{j=1,\ldots,10} KG_R(g^{\textrm{new}}_j;4000)[k] - \max_{j=1,\ldots,\textrm{length(DoE)}} \widehat R(g_j;4000) \Bigr)^+
\] 
for $k=1,\dots,1000$, that we represent in the histograms of Figure \ref{fig:histograms}. Additionally, we represent by the red line the \textit{a-posteriori} value of the expected improvement obtained on the new geometries: 
\[
\max_{i=1,\ldots,10} \widehat R(g^{\textrm{new}}_i;4000)-\max_{j=1,\ldots,\textrm{length(DoE)}} \widehat R(g_j;4000)
\]  
We call it the true value of the expected improvement.
Analogously, we consider also $1000$ realizations of $(KG_R(g^{\textrm{new}}_1;2500), \dots, KG_R(g^{\textrm{new}}_{10};2500))$ conditionally to the 300 estimated efficiencies at $Q=2500$ $m^3/h$.

\medskip 

\begin{figure} 
\begin{tabular}{cc}
\includegraphics[width=7cm,trim=0cm 0cm 0cm 0cm, clip]{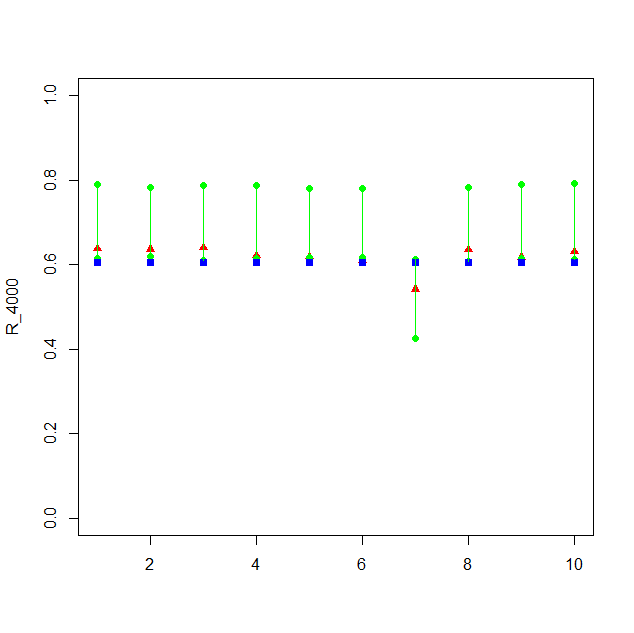}
&
\includegraphics[width=7cm,trim=0cm 0cm 0cm 0cm, clip]{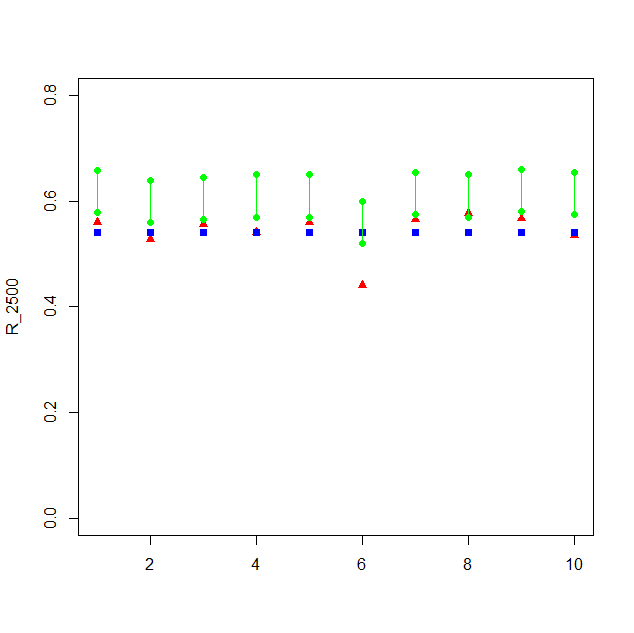}
\\
\end{tabular}

\caption{The results obtained on $\widehat R(\cdot{};4000)$ (respectively $\widehat R(\cdot{};2500)$) are represented in the left (resp. right). The red triangles give the values of the efficiency $R$ at $Q=4000$ for the 10 new geometries given by the expected improvement. The blue rectangles represents the current maximum. The green segments stand for the confidence intervals predicted by Kriging.}
\label{fig:new_geom_perf}
\end{figure}

In the left hand side of Figure \ref{fig:new_geom_perf}, we represent the values of the efficiency $R$ at $Q=4000$ $m^3/h$ for the 10 new geometries given by the expected improvement algorithm together with the value of the current maximum. This picture illustrates the fact that all the points correspond to exploitation (improvement of the promising regions) except the number 7 that corresponds to exploration. 
This fact can also be seen looking at the correlation matrix between the 10 new geometries presented in Table \ref{tab:matrix_corr}. The geometry numbered 7 is clearly uncorrelated from the others and corresponds to exploration. Moreover, notice that the confidence intervals given by Kriging are optimistic. 
Similarly in the right hand side of Figure \ref{fig:new_geom_perf}, we represent the values of the efficiency $R$ at $Q=2500$ $m^3/h$ for the 10 new geometries.

\begin{figure} 
\begin{tabular}{ccc}
\includegraphics{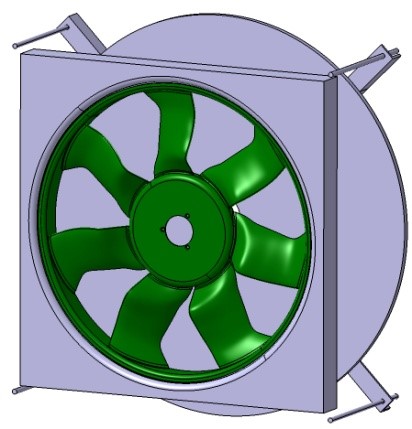}
&
\includegraphics{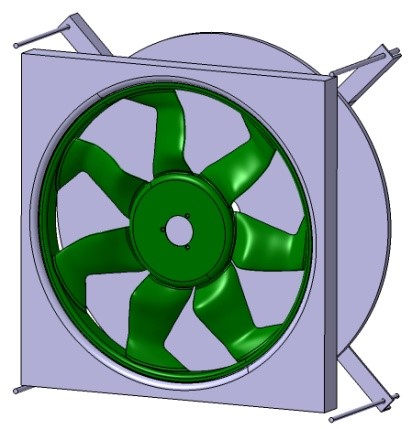}
&
\includegraphics{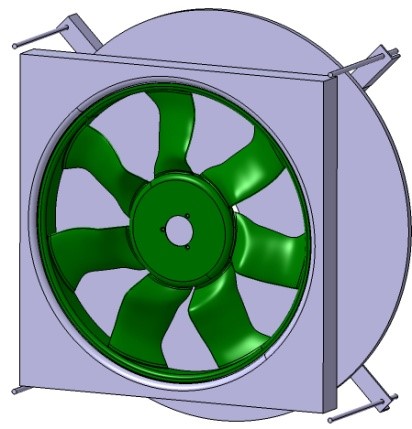}
\\
\includegraphics{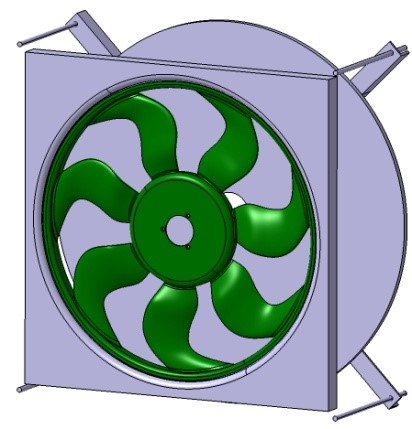}
&
\includegraphics{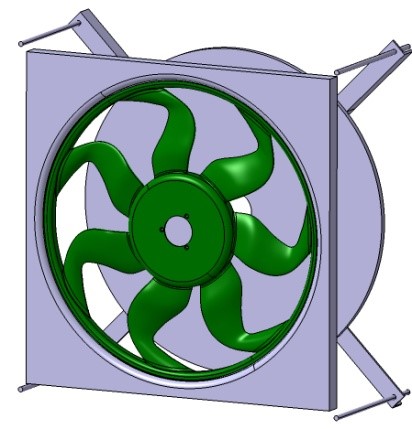}
&
\includegraphics{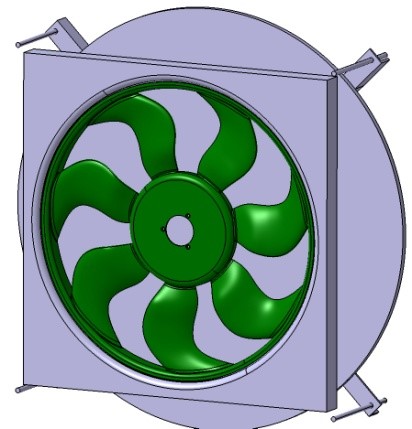}
\\
\end{tabular}

\caption{Illustration of the similarities with common ``genes''. Examples of optimized designs for high flow rate ($Q=4000m^3/h$ - top panel) and medium flow rate ($Q=2500m^3/h$ - bottom panel). }
\label{fig:fan_optim}
\end{figure}

Although the performances of the new geometries are finally at the bottom of the confident interval, it must be emphasized that they can be considered as very good design with high efficiencies (for this kind of ventilation system). This confirms that the tool is actually able to find the most interesting areas, with various solutions. As presented in Figure \ref{fig:fan_optim}, the targeted operating point for the optimization process determines some generic ``gene'' in the solutions. The searching method which is the NSGA II genetic algorithm, has obviously found for each of the optimization, either at high or a medium flow rate, two different sets of characteristics: at 4000$m^3/h$, the optimizer has selected straight blades with two discontinuities, respectively one close to the hub, and one close to the tip (top panel in Figure \ref{fig:fan_optim}). At 2500$m^3/h$, the design is being modified with a smoother shape from bottom to top and a backward blade sweep (meaning that the blade is curved in a direction opposite to the rotating one) (bottom panel in Figure \ref{fig:fan_optim}). 

Skilled engineers for this type of turbomachine have confirmed the relevancy of these observations, in particular the fact that backward sweep blades are good for efficiency and that straight ones are adapted to high flow rate. However, a so wide variety of design could not be obtained with classical iterative design methods, and the benefit of the meta-modeling is clearly its efficiency in proposing numerous designs in a short time frame.

\begin{table}
\begin{center}
\begin{tiny}
\begin{tabular}{rrrrrrr>{\columncolor{yellow}}rrrr}
     &       [,1]  &    [,2]    &   [,3]    &   [,4]    &   [,5]    &   [,6]       &[,7]    &   [,8]    &   [,9]   &   [,10]\\
$[1,]$ & 1.00000000 & 0.58156715 & 0.64818146 & 0.60873040 & 0.52468595 & 0.53664093 & 0.04283566 & 0.60534756 & 0.66981971 & 0.67070323\\
$[2,]$ & 0.58156715 & 1.00000000 & 0.53867748 & 0.59920321 & 0.59855007 & 0.60018872 & 0.03394723 & 0.50275974 & 0.58645696 & 0.58498154\\
$[3,]$ & 0.64818146 & 0.53867748 & 1.00000000 & 0.53393445 & 0.45723059 & 0.46774272 & 0.05433128 & 0.65886968 & 0.63105370 & 0.63127975\\
$[4,]$ & 0.60873040 & 0.59920321 & 0.53393445 & 1.00000000 & 0.59824959 & 0.60748421 & 0.03420089 & 0.47128678 & 0.62851412 & 0.63001910\\
$[5,]$ & 0.52468595 & 0.59855007 & 0.45723059 & 0.59824959 & 1.00000000 & 0.61347583 & 0.03068968 & 0.41150444 & 0.54010069 & 0.53917369\\
$[6,]$ & 0.53664093 & 0.60018872 & 0.46774272 & 0.60748421 & 0.61347583 & 1.00000000 & 0.03112121 & 0.41978435 & 0.55268651 & 0.55201157\\
\rowcolor{yellow} $[7,]$ & 0.04283566 & 0.03394723 & 0.05433128 & 0.03420089 & 0.03068968 & 0.03112121 & 1.00000000 & 0.05476946 & 0.03901714 & 0.03961607\\
$[8,]$ & 0.60534756 & 0.50275974 & 0.65886968 & 0.47128678 & 0.41150444 & 0.41978435 & 0.05476946 & 1.00000000 & 0.57885056 & 0.57779872\\
$[9,]$ & 0.66981971 & 0.58645696 & 0.63105370 & 0.62851412 & 0.54010069 & 0.55268651 & 0.03901714 & 0.57885056 & 1.00000000 & 0.67714934\\
$[10,]$ & 0.67070323 & 0.58498154 & 0.63127975 & 0.63001910 & 0.53917369 & 0.55201157 & 0.03961607 & 0.57779872 & 0.67714934 & 1.00000000\\
\end{tabular}
\end{tiny}
\end{center}
\caption{Conditional correlation matrix between the 10 new geometries considering $R(\cdot{};4000)$.}
\label{tab:matrix_corr}
\end{table}

\section{Conclusion}
\label{sec:concl}

A strategy for fan optimization, based on the use of an intense campaign of simulation has been proposed and tested. The method relies on a parametric model of the fan, which defines sets of parameters that are experimented in the design of experiment process. In order to address the difficulties related to the size of the domain in the dimension 15 and to the relative seldom runs, a meta-model based on a Kriging method has been built and further used to enrich the sampling. The combination of two tips has allowed improving the model: at first, a trend based on the turbomachine theory has been implemented for the efficiency in the Kriging model. Then several batches of additional runs have been proposed thanks to a criterion that seeks for the possible maximum improvement within the variance intervals. 
It has been observed that despite being too much optimistic, the results proposed by the genetic algorithm that interrogates the response surface are relevant and finally all have shown good efficiencies (except one over twenty). This good achievement indicates at first that the Kriging method is able to provide the good trends and can be used for optimization, in particular if the method can be improved on one hand by the use of turbomachine rules (here for instance using the efficiency as a trend), and on the other hand by a sequential strategy that exploits the expected improvement criterion. 
The proposed designs for a given targeted operating point have some similarities in their shapes, showing that the optimization process selects some characteristics which are deterministic. If it is on line with previous observations from the state of the art, it is remarkable how the method has provided so efficiently a wide variety of these best performer designs. 
All in all, the combination of design rules, numerical simulation and mathematics in meta-modeling is perceived as a very efficient method for optimization even in average dimensions. Perspectives are even more promising for the scientific community since the CPU cost is becoming every year more affordable and the pressure for optimized turbomachines in term of efficiency is high (due to economical and ecological concerns).

\medskip

\textbf{Acknowledgments.} The authors warmly thank the two anonymous reviewers for their helpful comments and suggestions that lead to a significant improvement of the manuscript. They also thank Fran\c cois Bachoc and C\'eline Helbert for their helpful discussions and suggestions. This work has been supported by the French National
Research Agency (ANR) through PEPITO project 
(no ANR-14-CE23-0011).

\bibliographystyle{abbrv}
\bibliography{biblio_EI}
\end{document}